%% file: main.tex
\documentclass[abc,preprint,nocopyrightspace]{vldb}

\usepackage{courier}
\usepackage[scaled]{helvet}
\usepackage{url}
\usepackage{listings}
\usepackage{enumitem}
\usepackage{algorithm}
\usepackage{graphicx}
\usepackage{comment}
\usepackage{amsmath}
\usepackage{amssymb}
\usepackage{bbm, dsfont}
\usepackage{mathrsfs}
\usepackage{xcolor}
\usepackage{tikz}
\usepackage{pgfplots}
\usepackage{esvect}
\usepackage{xspace}
\usepackage{array, multirow, multicol}
\usepackage{rotating, makecell} 
\usepackage[noend]{algpseudocode}
\usepackage{balance}
\usepackage{mathpartir}
\usepackage{stmaryrd}
\usepackage{subfigure}
\usepackage{mhchem}

\newcommand{\system}{{\sc Sqlizer}\xspace}
\newcommand{\nalir}{{\sc Nalir}\xspace}

\newcommand{\join}[4]{#1 \ce{_{\it #2}\hspace{-0.03in}\bowtie\hspace{-0.04in}_{\it #3}} #4}

\newcommand*{\comments}{} 
\ifdefined\comments
\newcommand{\navid}[1]{\textcolor{blue}{\textbf{NAVID:} #1}}
\newcommand{\yuepeng}[1]{\textcolor{magenta}{\textbf{YUEPENG:} #1}}
\newcommand{\isil}[1]{\textcolor{red}{\textbf{I\c{S}IL:} #1}}
\newcommand{\todo}[1]{\textcolor{red}{#1}}
\else
\newcommand{\navid}[1]{}
\newcommand{\yuepeng}[1]{}
\newcommand{\isil}[1]{}
\newcommand{\todo}[1]{}
\fi

\newtheorem{example}{Example}

\newcommand{\irule}[2]%
   {\mkern-2mu\displaystyle\frac{#1}{\vphantom{,}#2}\mkern-2mu}

\newcommand*{\termdelim}{~~|~~}

\newcommand*{\vdashs}{\vdash_{\hspace{-0.03in}s}}

\algnewcommand{\IIf}[2]{\State\algorithmicif\ #1\ \algorithmicthen\ #2\ }
\algnewcommand{\IElseIf}[2]{\State\algorithmicelse\ \algorithmicif\ #1\ \algorithmicthen\ #2\ }
\algnewcommand{\IElse}[1]{\State\algorithmicelse\ #1\ }
\newcommand*{\query}{\mathcal{Q}}
\newcommand*{\sketch}{\mathcal{S}}
\newcommand*{\inst}{\mathcal{I}}
\newcommand*{\prob}{\mathcal{P}}
\newcommand*{\res}{\mathcal{R}}

\makeatletter
\def\@copyrightspace{\relax}
\makeatother
\begin{document}


\title{Type- and Content-Driven Synthesis of \\ SQL Queries from Natural Language}


%
%
%
%

\numberofauthors{1}

\author{
\alignauthor
Navid Yaghmazadeh ~~~ Yuepeng Wang ~~~ Isil Dillig ~~~ Thomas Dillig\\
University of Texas at Austin, USA\\
\{nyaghma, ypwang, isil, tdillig\}@cs.utexas.edu
}

\maketitle

\input{abstract}

\input{intro}

\input{overview}
\input{prelim}

\input{generation}

\input{completion}

\input{refinement}

\input{impl}

\input{evaluation}
\input{related}

\input{conclusion}

\bibliographystyle{abbrv}
\bibliography{main}

\balance

\end{document}

%% file: abstract.tex
\begin{abstract}
This paper presents a new technique for automatically synthesizing SQL queries from natural language. Our technique is fully automated, works for \emph{any} database without requiring additional customization, and  does not require users to know  the underlying database schema.
Our method achieves these goals by combining natural language processing, program synthesis, and automated program repair. Given the user's English description, our technique first uses semantic parsing to generate a \emph{query sketch}, which is subsequently completed using type-directed program synthesis and assigned a confidence score using database  contents. However, since the user's  description may not accurately reflect the actual database schema, our approach also performs \emph{fault localization} and repairs the erroneous part of the sketch. This \emph{synthesize-repair} loop is repeated until the algorithm infers a query with a sufficiently high confidence score. We have implemented the proposed technique  in a tool called \system and evaluate it on three different databases. Our experiments show that the desired query is ranked within the top 5 candidates in close to 90\% of the cases.

\end{abstract}

%% file: intro.tex
\section{Introduction}\label{sec:intro}

With the rapid proliferation of data in today's information-rich world, many people across a variety of professions need to query data stored in some kind of relational database. While relational algebra and its incarnations in modern programming languages, such as SQL, JPQL, and Linq, provide a rich and flexible mechanism for querying data, it is often non-trivial for end-users to ``program'' in these languages:

\begin{itemize}
\item First, in order to successfully retrieve the desired data, the user must be aware of the underlying data representation. In particular, the user must know about specific database tables and their attributes, which is often not a realistic assumption.
\item Second, despite being declarative and easier to master than most other programming languages, there is still a craft involved in ``translating" database queries to relational algebra. Hence,  most newcomers require significant practice to write non-trivial SQL queries that involve join operations and subqueries.
\end{itemize}

In this paper, we propose a new technique for automatically synthesizing relational algebra expressions from natural language. 
While there has been considerable interest in \emph{natural language interfaces to databases (NLIDB)} over the last several decades~\cite{nlidb-survey1,nalir,precise,precise2,nalix},  existing techniques suffer from various shortcomings. For example,  many techniques (e.g., ~\cite{mooney1,mooney3}) require the system to be trained on a specific database and are therefore not \emph{database-agnostic} (i.e., require customization for each database). Other techniques (e.g.,~\cite{precise,nalir,nalix}) aim to work on a broader class of databases but suffer from other shortcomings. For instance, the {\sc Precise} system~\cite{precise} only works for a subset of natural language descriptions, and {\sc Nalir}~\cite{nalir} requires guidance from the user to choose the right query structure and intended table/column names for it to be effective.

\begin{figure*}
\begin{center}
\includegraphics[scale=0.24]{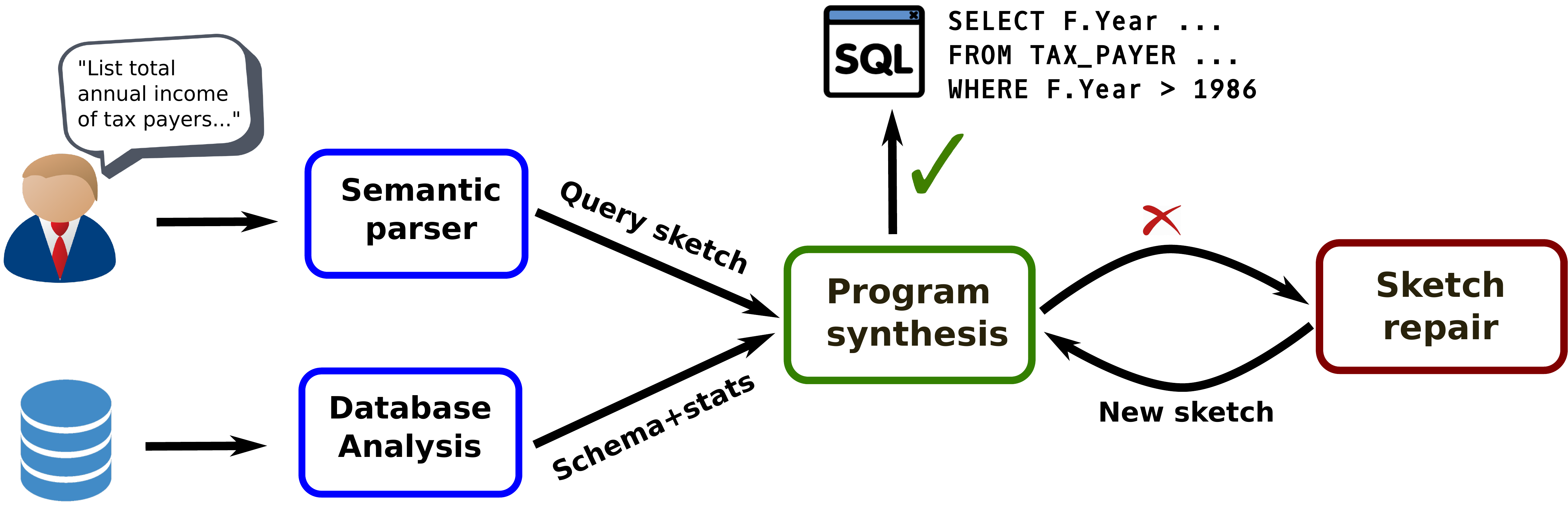}
\end{center}
\vspace{-0.2in}
\caption{Schematic overview of our approach}\label{fig:overview}
\end{figure*}

In this paper, we describe a new approach, and its implementation in a tool called \system, that addresses many of the shortcomings of existing NLIDB systems. Specifically, our method is  database-agnostic, fully automated, and does not impose restrictions on the class of natural language descriptions that can be provided by the user. As we show experimentally, \system achieves high precision across multiple different databases and significantly outperforms {\sc Nalir}, a state-of-the-art NLIDB system~\cite{nalir}, when {\sc Nalir} is used in its non-interactive setting.

As shown schematically in Figure~\ref{fig:overview}, the key insight underlying our method is to combine semantic parsing techniques from the NLP community with type-directed program synthesis and automated program repair:

\vspace{0.05in}
\noindent
{\bf \emph{Semantic parsing}}. At a high level, our method uses semantic parsing~\cite{mooney2,sem-parse1,sem-parse2}  to translate the user's English description into a so-called \emph{query sketch (skeleton)}. Since a query sketch only specifies the shape -- rather than the full content -- of the query (e.g., join followed by selection followed by projection), the semantic parser does not need to know about the names of database tables/columns. Hence, the use of query sketches (as opposed to full-fledged SQL queries) allows  our semantic parser to effectively  translate the English description into a suitable formal representation without requiring any database-specific training.

\vspace{0.05in}
\noindent
{\bf \emph{Type-directed synthesis}}. Once a query skeleton is generated, our technique employs type-directed program synthesis to complete the sketch. In particular, \system enumerates  well-typed completions of the query skeleton with the aid of the underlying database schema.  However, because there are  typically {many}  well-typed terms, our approach assigns a \emph{confidence score} to each possible completion of the sketch. Our synthesis algorithm uses both the  \emph{contents} of the database as well as natural language hints embedded in the  sketch when assigning confidence scores to SQL queries.

\vspace{0.05in}
\noindent
{\bf \emph{Sketch repair}}.
Since our approach  assumes that users may not be familiar with the underlying data organization, the initial query sketches generated using semantic parsing may not accurately reflect the structure of the target query. Hence, there may not be any well-typed, high-confidence completions of the sketch.  For example, consider a scenario in which the user believes that the desired data is stored in a single database table, although it in fact requires joining two different tables. Since the user's English  description does not adequately capture the structure of the desired query, the initial query sketch needs to be repaired. 
 Our approach deals with this challenge by (a) performing fault localization to pinpoint the likely cause of the error, and (b) using a database of ``repair tactics" to refine the initial sketch.

 \vspace{0.05in}
\noindent
{\bf \emph{System architecture}}. 
The overall architecture of \system is shown schematically in Figure~\ref{fig:overview} and is a based on a refinement loop that  incorporates each of the three key ideas outlined above. Given the user's natural language description, \system first generates the top $k$ most likely query sketches $Q$ using semantic parsing, and, for each query skeleton $q \in Q$,  it tries to synthesize a well-typed, high-confidence completion of $q$.  If no such completions can be found, \system tries to identify the root cause of failure and automatically repairs the suspect parts of the sketch. Once \system enumerates all high-confidence queries that can be obtained using at most $n$ repairs on $q$, it then moves on to the next most-likely query sketch. At the end of this \emph{parse-synthesize-repair} loop, \system ranks all queries based on their  confidence scores and presents the top $m$ results to the user.

 \vspace{0.05in}
\noindent
{\bf \emph{Results}}.
We have evaluated  \system on 455 queries involving three different databases, namely MAS (Microsoft Academic Search), IMDB, and Yelp. Our evaluation shows that the desired query is ranked within  \system's top 5 solutions approximately 90\% of the time and within top 1 close to 80\% of the time.  We also compare \system against a  state-of-the-art NLIDB tool,   \nalir, and show that \system performs significantly better across all three databases.

 \vspace{0.05in}
\noindent
{\bf \emph{Contributions}}.
In summary, this paper makes the following key contributions:

\begin{itemize}
\item We present \system, an end-to-end system for synthesizing SQL queries from natural language. \system is fully automated, database-agnostic, and does not require  users to know the underlying database schema.
\item  We design and implement a semantic parser to translate natural language descriptions into query skeletons.
\item We describe a new type- and content-directed synthesis methodology for obtaining SQL queries from  query sketches.
\item We show how to refine query sketches using automated fault localization and repair tactics.
\item We  evaluate \system on a set of 455 queries involving three databases. Our results show that  the desired query is ranked number one in 78.4\% of benchmarks and within the top 5 in 88.3\% of the benchmarks.
\end{itemize}

 \vspace{0.05in}
\noindent
{\bf \emph{Organization}}. The rest of the paper is organized as follows: We first provide an overview of our approach through a simple motivating example (Section~\ref{sec:overview}). We then provide some background on an extended version of relational algebra in Section~\ref{sec:prelim}, and  give a brief overview of \system's semantic parser in Section~\ref{sec:parse}. Our main technical contributions are described in Section~\ref{sec:completion} (type- and content-directed sketch completion) and Section~\ref{sec:repair} (fault localization and repair), followed by a brief discussion of our implementation in Section~\ref{sec:impl}. Finally, we present the results of our empirical evaluation in Section~\ref{sec:eval} and survey related work in Section~\ref{sec:related}.

%% file: overview.tex
\section{Overview}\label{sec:overview}

\begin{figure*}
\begin{center}
\includegraphics[scale=0.55]{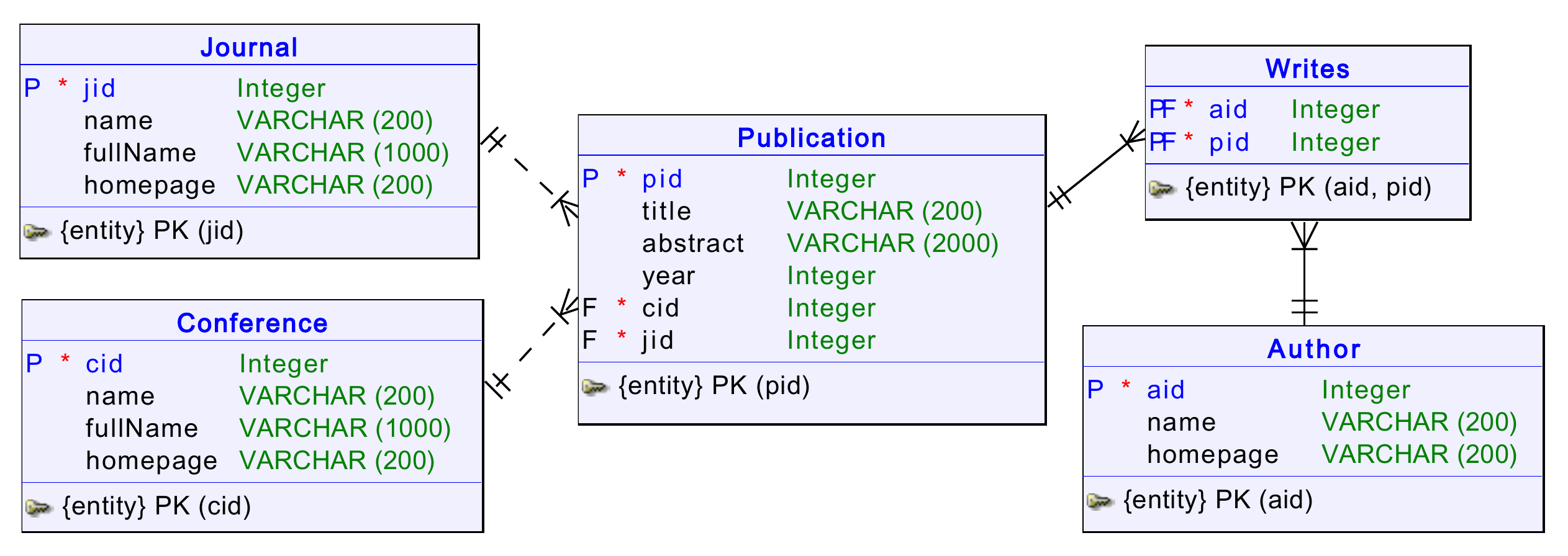}
\end{center}
\vspace{-0.25in}
\caption{Simplified schema for the Microsoft Academic Search (MAS) database}\label{fig:schema}
\end{figure*}

In this section, we give a high-level overview of our technique with the aid of a simple motivating example.  Figure~\ref{fig:schema} shows the relevant portion of the schema for the Microsoft Academic Search (MAS) database, and suppose that we would like to synthesize a database query to retrieve the number of papers in 
VLDB 2010. To use our tool, the user provides an English description, such as \emph{``Find the number of papers in VLDB 2010"}. We now outline the steps taken by \system in synthesizing the desired SQL query.

\vspace{0.1in}\noindent
{\bf \emph{Sketch generation.}} Our approach first uses a semantic parser to generate the top $k$ most-likely program sketches. For this example, the highest-ranked query sketch returned by the semantic parser is the following~\footnote{\scriptsize We actually represent query sketches using an extended version of relational algebra. However,  in this section, we present query skeletons using SQL for easier readability.}:

\begin{verbatim}
SELECT count(?[papers]) FROM ??[papers] 
WHERE ? = "VLDB 2010"
\end{verbatim}
Here,  {\tt ??} represents an unknown table, and {\tt ?}'s represent unknown columns. Where present, the words written in square brackets represent so-called ``hints" for the corresponding hole. For example, the second hint in this  sketch indicates that the table represented by {\tt ??} is semantically similar to the English word ``papers".

\vspace{0.1in}\noindent
{\bf \emph{First iteration.}}  Starting from the above sketch $\sketch$, \system enumerates all well-typed completions $q_i$ of $\sketch$, together with a score for each $q_i$. In this case, there are many possible well-typed completions of $\sketch$, however, none of the $q_i$'s meet our confidence threshold. For instance, one of the reasons why \system fails to find a high-confidence query is that there is no entry called ``VLDB 2010" in any of the database tables.

Next, \system performs fault localization on $\sketch$ to identify the root cause of failure (i.e., not meeting confidence threshold). In this case, we determine that the likely root cause is the predicate {\tt ? = "VLDB 2010"} since there is no database entry matching ``VLDB 2010", and our synthesis algorithm has assigned a low confidence score to this term. Next, we repair the sketch by splitting the where clause into two separate conjuncts. As a result, we obtain the following refinement $\sketch'$ of the initial sketch $\sketch$:

\begin{verbatim}
SELECT count(?[papers]) FROM ??[papers] 
WHERE ? = "VLDB" AND ? = 2010
\end{verbatim}

\vspace{0.1in}\noindent
{\bf \emph{Second iteration.}} Next, \system tries to complete the refined sketch $\sketch'$ but it again fails to find a high-confidence completion of $\sketch'$. In this case, the problem is that there is no single database table that contains both the entry ``VLDB" as well as the entry ``2010".  Going back to fault localization, we now determine that the most likely problem  is the term {\tt ??[papers]}, and we try to repair it by introducing a join. 
As a result, the new sketch $\sketch''$ now becomes:

\begin{verbatim}
SELECT count(?[papers]) 
FROM  ??[papers] JOIN  ?? ON ? = ?
WHERE ? = "VLDB" AND ? = 2010
\end{verbatim}

\vspace{0.1in}\noindent
{\bf \emph{Third iteration.}}
After going back to the sketch completion phase a third time, we are now able to find a high-confidence instantiation $q$ of $\sketch''$. In this case, the highest ranked completion of $\sketch''$ corresponds to the following query:

\begin{verbatim}
SELECT count(Publication.pid) 
FROM  Publication JOIN  Conference 
ON Publication.cid = Conference.cid
WHERE Conference.name = "VLDB" 
      AND Publication.year = 2010
\end{verbatim}

This query is indeed the correct one, and running it on the MAS database yields the number of papers in VLDB 2010. 

%% file: prelim.tex
\section{Preliminaries}\label{sec:prelim}

\input{symb-sql}

Figure~\ref{fig:symb-sql} presents a variant of relational algebra that we use to formalize our synthesis approach throughout this paper. As shown in Figure~\ref{fig:symb-sql},  \emph{relations}, denoted as $T$ in the grammar, include \emph{tables} $t$ stored in the database or  \emph{views} obtained by applying the relational algebra operators,  projection ($\Pi$), selection ($\sigma$), and join ($\bowtie$). As standard, projection $\Pi_L(T)$ takes a relation $T$ and a column list $L$  and returns a new relation that only contains the columns in  $L$. The selection operation $\sigma_\phi(T)$ yields a new relation that only contains rows satisfying $\phi$ in $T$. The join operation $\join{T_1}{c_1}{c_2}{T_2}$ composes two relations $T_1$, $T_2$ such that the result   contains exactly those rows of $T_1 \times T_2$  satisfying $c_1 = c_2$, where $c_1, c_2$ are columns in $T_1, T_2$ respectively.

In this paper, we make a number of assumptions that simplify our technical presentation. First, we assume that every column in the database has a unique name. Note that we can easily enforce this restriction in practice by appending the table name to each column name. Second, we only consider \emph{equi-joins} because they are the most commonly used join operator, and it is easy to extend our techniques to other kinds of join operators (e.g., $\theta$-join). However, please observe that the grammar from Figure~\ref{fig:symb-sql} allows nested queries. For instance, selections can occur within other selections and joins as well as inside predicates $\phi$.

Unlike standard relational algebra,  the relational algebra variant shown in Figure~\ref{fig:symb-sql} also allows aggregate functions as well as a group-by operator. For conciseness, aggregate functions  $f \in \emph{AggrFunc} = \emph{\{max, min, avg, sum, count\}}$ are specified as a subscript in the projection operation. In particular,   $\Pi_{f(c)}(T)$ yields a single aggregate value obtained by applying $f$ to column $c$ of relation $T$.  Similarly, group-by operations are also specified as a subscript in the projection operator. Specifically, $\Pi_{g(f(c_1), c_2)}(T)$ divides rows of $T$ into groups $g_i$ based on values stored in column $c_2$ and, for each $g_i$, it yields the aggregate value $f(c_1)$.

\begin{example}
\begin{figure}[t]
\centering
\subfigure[Grades]{
\begin{tabular}{| c | c | c | c |}
\hline            
    id & name & score & {cid\_fk} \\
    \hline            
    1 & John & 60 & 101 \\
    \hline            
    2 & Jack & 80 & 102 \\
    \hline            
    3 & Jane & 80 & 103 \\
    \hline            
    4 & Mike & 90 & 104 \\
    \hline            
    5 & Peter & 100 & 103 \\
    \hline            
    6 & Alice & 100 & 104 \\
    \hline  
\end{tabular}
}
\subfigure[Courses]{
\begin{tabular}{| c | c | c |}
\hline            
    cid & cname & dept \\
    \hline            
    101 & C1 & CS \\
    \hline            
    102 & C2 & EE \\
    \hline            
    103 & C3 & CS \\
    \hline            
    104 & C4 & EE \\
    \hline  
\end{tabular}
}
\vspace{-0.1in}
\caption{Example tables} \label{fig:ex}
\end{figure}

Consider  the ``Grades'' and ``Courses'' tables from Figure~\ref{fig:ex}, where column names with suffix ``\_fk" indicate foreign keys. Here, $\Pi_\emph{avg(score)}(\emph{Grades})$ evaluates to 85, and {$\Pi_\emph{g(avg(score), dept)}({\join{\emph{Grades}}{cid\_fk}{cid}{\emph{Courses}}})$} yields the following table:
\begin{center}
\begin{tabular}{| c | c |}
\hline            
    dept & avg(score) \\
    \hline            
    CS & 80 \\
    \hline            
    EE & 90 \\
    \hline  
\end{tabular}
\end{center}

To provide an example of nested queries, suppose that a user wants to retrieve all students with the highest score. We can express this query as: \[
\Pi_\emph{name}(\sigma_{\emph{score}=\Pi_\emph{max(score)}(\emph{Grades})}(\emph{Grades}))\]
For the tables from Figure~\ref{fig:ex}, this query yields a table with two rows, Peter and Alice.
\end{example}

%% file: symb-sql.tex
\begin{figure}
\centering
\[
\begin{array}{r c l}
    T &:=& \Pi_L(T) \termdelim \sigma_\phi(T) \termdelim \join{T}{c}{c}{T} \termdelim t \\
    L &:=& L, L \termdelim c \termdelim f(c) \termdelim g(f(c), c) \\
    E &:=& T \termdelim c \termdelim v \\
    \phi &:=& \phi~lop~\phi \termdelim \neg \phi \termdelim c~op~E \\
    op &:=& \leq \termdelim < \termdelim = \termdelim > \termdelim \geq \\
    lop &:=& \land \termdelim \lor \\
\end{array}
\]
\vspace{-0.2in}
\caption{Grammar of Extended Relational Algebra. Here, $t, c$ denote table and column names; $f$ denotes an aggregate function, and $v$ denotes a value.}
\label{fig:symb-sql}
\end{figure}

%% file: generation.tex
\section{Sketch Generation}\label{sec:parse}
In this section, we give a brief overview of \system's semantic parser.

\vspace{0.1in} \noindent
{\bf \emph{Background on semantic parsing.}}
The goal of a semantic parser is to map a phrase in  natural language to a so-called \emph{logical form} which represents its meaning.  A logical form is an unambiguous artificial language, typically specified using a context-free grammar. Previous work on semantic parsing has used a variety of different logical form representations,  including lambda calculi~\cite{carpenter}, database query languages~\cite{mooney1}, and natural logics~\cite{manning}.

Due to the ambiguity of natural language, an English sentence is typically associated with multiple different logical forms. A key idea underlying modern semantic parsers is to use statistical methods (i.e., machine learning) to predict the most likely logical form for a given utterance. Given a set of training data consisting of pairs of English sentences and their corresponding logical form, semantic parsers typically train a statistical model to predict the likelihood that a given English sentence is mapped to a particular logical form. Hence, the output of a semantic parser is a list of logical forms $x_i$, each associated with a probability that the English sentence corresponds to $x_i$.

\vspace{0.1in} \noindent
{\bf \emph{\system's Semantic Parser.}} The logical forms used in \system take the form of \emph{query skeletons}, which are produced according to the grammar from Figure~\ref{fig:symb-sketch}. Intuitively, a query skeleton is a relational algebra term with missing table and column names. In this work, we use query skeletons  as the underlying logical form representation because it is infeasible to accurately map utterances to full SQL queries \emph{without training on a specific database}. In other words, the use of query skeletons allows us to map English sentences to logical forms in a \emph{database-agnostic manner}.

\input{symb-sketch}

Let us now consider the query skeletons $\chi$ from Figure~\ref{fig:symb-sketch} in more detail. Here, $?h$ represents an unknown column with \emph{hint} $h$, which is just a natural language description of the unknown. Similarly, $??h$ represents an unknown table name with corresponding hint $h$. If there is no hint associated with a hole, we simply write $?$ for columns and $??$ for tables.

To map English sentences to query skeletons, we have implemented our own semantic parser on top of the 
{\sc Sempre} framework~\cite{sempre}, which is a toolkit for training semantic parsers. Given an utterance $u$, our parser generates all possible query skeletons $\sketch_i$  and assigns each $\sketch_i$  a score that indicates the likelihood that $\sketch_i$ is the intended interpretation of $u$. This score is calculated based on a set of pre-defined features. More precisely, given an utterance $u$ and weight vector $w$, the parser maps each query skeleton $\sketch_i$ to a $d$-dimensional feature vector $\phi(u, \sketch_i) \in \mathbb{R}^d$ and computes the likelihood score for  $\sketch_i$ as the weighted sum of its features: 
$$score(u, \sketch_i) \ = \ \vec{w} \ . \ \phi(u, \sketch_i) \ = \ \sum_{j=1}^{d} w_j \cdot {\phi(u, \sketch_i)}_j$$
Since our semantic parser inherits the features pre-defined in {\sc Sempre}, we do not describe them in detail and refer the interested reader to previous work on this topic~\cite{sem-parse1}.

As standard, our semantic parser performs training to optimize the weight vector $\vec{w}$ such that the scoring function associates a higher value with the intended logical form of an utterance (see Section~\ref{sec:impl} for more details). Since we also inherit {\sc Sempre}'s existing learning algorithm based on \emph{stochastic gradient descent (SGD)}, we refer the reader to ~\cite{sem-parse1} for more details about learning.

%% file: symb-sketch.tex
\begin{figure}
\centering
\[
\begin{array}{r c l}
    \chi &:=& \Pi_\kappa(\chi) \termdelim \sigma_\psi(\chi) \termdelim \join{\chi}{?h}{?h}{\chi} \termdelim ??h \\
    \kappa &:=& \kappa,  \kappa \termdelim ?h \termdelim f(?h) \termdelim g(f(?h), ?h) \\
    \eta &:=& \chi \termdelim ?h \termdelim v \\
    \psi &:=& \psi~lop~\psi \termdelim \neg \psi \termdelim ?h~op~\eta \\
    op &:=& \leq \termdelim < \termdelim = \termdelim > \termdelim \geq \\
    lop &:=& \land \termdelim \lor \\
\end{array}
\]
\vspace{-0.2in}
\caption{Sketch Grammar. Here, $v$ denotes a value, $h$ represents a natural language hint, and $f$ is an aggregate function.}
\label{fig:symb-sketch}
\end{figure}

%% file: completion.tex
\section{Sketch Completion} \label{sec:completion}

Given a query sketch generated by the semantic parser, \system tries to complete this sketch by instantiating the holes with  concrete table and column names defined in the database schema. The sketch completion procedure is \emph{type-directed} and treats each database table as a record type $\{ (c_1: \beta_1), \ldots (c_n: \beta_n) \}$, where $c_i$ is a column name and $\beta_i$ is the type  of the values stored in column $c_i$. While our sketch completion algorithm only generates well-typed terms, there are typically many possible well-typed completions of the sketch. As mentioned in Section~\ref{sec:intro}, our algorithm overcomes this challenge by associating a confidence score with each completion. In particular, our scoring function utilizes two valuable sources of information:

\begin{itemize}
\item \emph{Natural language hints:} 
Our sketch completion engine computes a similarity metric between the hints embedded in the query sketch and the  names of the database tables and columns. This similarity metric is taken into account when assigning a score to each possible completion of the sketch.

\item \emph{Database contents:} Our approach  also uses the contents of the database when assigning  scores to queries. To gain intuition about why database contents are useful, consider a candidate term $\sigma_\phi(T)$.  If there are no entries in relation $T$ satisfying predicate $\phi$, then $\sigma_\phi(T)$ has a low probability of occurring in the target query.
\end{itemize}

Keeping the above intuition in mind, we describe our sketch completion algorithm using the inference rules shown in Figures~\ref{fig:rules-view} and~\ref{fig:rules-spec}. Specifically, the top-level sketch completion rules derive  judgments of the form
$
\Gamma \vdash \chi \Downarrow T: \tau, p
$ where $\Gamma$ is a type environment mapping each table $t$ in the database to its corresponding type. The meaning of this judgment is that sketch $\chi$ instantiates to a relational algebra term $T$  of type $\tau$ with ``probability" $p \in [0, 1]$. The higher the score $p$, the more likely it is that $T$ is a valid completion of sketch~$\chi$.

Figure~\ref{fig:rules-spec} presents the helper rules used for instantiating so-called \emph{specifiers}. A specifier $\omega$ of a sketch $\chi$ is any subterm of $\chi$ that does not correspond to a relation. For example, the specifier for  $\Pi_\kappa(\chi)$ is $\kappa$, and the specifier for $\sigma_\psi(\chi)$ is $\psi$. The instantiation rules for specifiers are shown in Figure~\ref{fig:rules-spec} using judgements of the form:
\[ \Gamma, \tau \vdashs \omega \Downarrow Z : \tau', p
\]
Here, type $\tau$ used on the left-hand side of the judgment denotes the type of the table that $\omega$ is supposed to qualify. For instance, if the parent term of $\omega$ is $\Pi_\omega(\chi)$, then $\tau$ represents the type of relation $\chi$. Hence, the meaning of this judgment is that, under the assumption that $\omega$ qualifies a relation of type $\tau$, then $\omega$ instantiates to a term $Z$ of type $\tau'$ with probability~$p$.

\vspace{0.1in}
\noindent
{\bf \emph{Instantiating Relation Sketches.}} Consider a sketch of the form $??h$ indicating an unknown database table with hint $h$. In principle, we can instantiate $??$  with any table $t$ in the database (i.e., $t$ is in the domain of $\Gamma$). However, as shown in the first rule from Figure~\ref{fig:rules-view}, our approach uses the hint $h$ to compute the likelihood that $t$ is the intended completion of $??$. Specifically, we use the $sim$ procedure to compute a similarity score between hint $h$ and table name $t$ using Word2Vec~\cite{word2vec}, which uses a two-layer neural net to group similar words together in vector-space.~\footnote{\scriptsize Our use of Word2Vec is similar to the use of WordNet in \nalir. Specifically, \nalir computes a similarity score (using WordNet) between English words in their dependency parse tree and database elements in the query tree.  Hence, both \system and \nalir are predicated on the intuition that the names of schema elements are typically human-readable and meaningful. 
}

\input{symb-synth}

\input{rules-view}

Let us now consider the completion of query sketches that involve projection and selection (rules Proj, Sel in Figure ~\ref{fig:rules-view}). Given a sketch of the form $\Pi_\kappa(\chi)$ (resp. $\sigma_\psi(\chi)$), we first recursively instantiate the sub-relation $\chi$ to $T$. Now, observe that the columns used in  specifiers $\kappa$ and $\psi$ can only refer to columns in $T$; hence, we use the type $\tau$ of $T$ when instantiating specifiers $\kappa, \psi$. Now, assuming  $\chi$ instantiates to $T$ with probability $p_1$ and $\kappa$ (resp. $\psi$) instantiates to $L$ (resp. $\phi$) with probability $p_2$, we need to combine  $p_1$ and $p_2$ to determine a score for $\Pi_L(T)$ (resp. $\sigma_\phi(T)$). Towards this goal, we define an operator $\otimes$ that is used to compose different scores. While there are many possible ways to compose scores, our implementation defines $\otimes$ as the geometric mean of $p_1, \ldots, p_n$:
\[
p_1 \otimes ... \otimes p_n = \sqrt[n]{p_1 \times ... \times p_n}
\]

Observe that our definition of $\otimes$ is not the standard way to combine probabilities (i.e., multiplication). We have found that this definition of $\otimes$ works better in practice, as it does not penalize long queries and allows a more meaningful comparison between different-length queries. However, one implication of this choice is that the scores of all possible completions do not add up to 1. Hence, these scores are not true ``probabilities'' in the technical sense, but, with slight abuse of terminology, we nevertheless use the terms score and probability interchangably throughout the paper.

Finally, let us consider the \emph{Join} rule for completing sketches of the form $\join{\chi_1}{?h_1}{?h_2}{\chi_2}$. As before, we first complete the nested sketches $\chi_1$ and $\chi_2$, and then instantiate $?h_1$ and $?h_2$ under the assumption that $\chi_1, \chi_2$ have types $\tau_1, \tau_2$ respectively. 
The function $P_{\bowtie}(c_1, c_2)$ used in the \emph{Join} rule assigns a higher score to term $\join{T_1}{c_1}{c_2}{T_2}$ if  column $c_1$ is a foreign key referring to column $c_2$ in table $T_2$ (or vice versa). Intuitively, if $c_1$ in table $T_1$ is a foreign key referring to $c_2$ in Table $T_2$, then there is a high probability that the term $\join{T_1}{c_1}{c_2}{T_2}$ appears in the target query.

\vspace{0.1in}\noindent
{\bf \emph{Instantiating Specifiers}}. In our discussion so far, we ignored how to instantiate and assign confidence scores to specifiers. 
 Let us now turn our attention to the rules from Figure~\ref{fig:rules-spec} for addressing this problem.
 
 In the simplest case, consider instantiating a column of the form $?h$. As shown in the \emph{Col} rule of Figure~\ref{fig:rules-spec},  we ensure that the candidate instantiation $c$ is actually a column of the table associated with the parent relation. This is exactly the reason why the inference rules for specifiers require the type $\tau$ of the parent in addition to environment $\Gamma$. In other words, our sketch completion algorithm \emph{stratifies} its enumerative search based on types and always fills holes of type relation before moving on to other types.

Since most of the rules in Figure~\ref{fig:rules-spec} are quite similar to the ones from Figure~\ref{fig:rules-view}, we do not explain  them in detail. However, we would like to draw the reader's attention to the \emph{Pred} rule for instantiating predicate sketches of the form $?h \ \emph{op} \ \eta$. Recall that a predicate $c \ \emph{op} \ v$ evaluates to true for exactly those values $v'$ in column $c$ for which the predicate $v' \ \emph{op} \ v$ is true. Now, if $c$ does not contain any entries $v'$ for which the $v' \ \emph{op} \ v$ evaluates to true, there is a low, albeit  non-zero, probability that $c \ \emph{op} \ v$ is the intended predicate. To capture this intuition, the \emph{Pred} rule uses the following $\emph{P}_{\phi}$ function when assigning a confidence score:

\input{rules-spec}

\[
    \emph{P}_{\phi}(c~op~E) =
    \begin{cases}
    1 - \epsilon & \text{if}~\exists v' \in \emph{contents}(c).~v'~op~E = \top\\
    \epsilon & \text{otherwise}
    \end{cases}
\]

Here, $\epsilon$ is a small, non-zero constant that indicates low confidence. Hence, predicate $c~\emph{op}~E$ is assigned a low score if there are no entries satisfying it in the database.

\begin{example}
 Consider again the (tiny) database  shown in Figure~\ref{fig:ex} and the following query sketch:
\[
    \Pi_{\emph{g(avg(?score), ?department)}}(\join{??\emph{score}}{?}{?}{??})
\]
According to the rules from Figures~\ref{fig:rules-view} and \ref{fig:rules-spec}, the query
\[
    \Pi_{\emph{g(avg(score), dept)}}(\join{\emph{Grades}}{cid\_fk}{cid\_fk}{\emph{Grades}})
\]
is not a valid completion  because it is not ``well-typed" (i.e., {\tt dept} is not a column in $\join{\emph{Grades}}{cid\_fk}{cid\_fk}{\emph{Grades}}$). In contrast, the query 
\[
    \Pi_{\emph{g(avg(score), dept)}}(\join{\emph{Grades}}{id}{cid}{\emph{Courses}})
\]
is well-typed but is assigned a low score because  the column {\tt id} in Grades is not a foreign key referring to column {\tt cid} in  Courses. As a final example, consider the query:
\[
    \Pi_{\emph{g(avg(score), dept)}}(\join{\emph{Grades}}{cid\_fk}{cid}{\emph{Courses}})
\]
This query is both well-typed and is assigned a high score close to $1$.
\end{example}

%% file: symb-synth.tex
\begin{figure}
\centering
\[
\begin{array}{r c l}
    \beta &:=& \emph{Number}~\mathbb{N} \termdelim \emph{Bool}~\mathbb{B} \termdelim \emph{String}~\mathbb{S} \\
    \tau & := & \beta \ | \ \{(c_1: \beta_1), \cdots, (c_n: \beta_n)\} \\
    \Gamma &::& \emph{Table} \to \tau \\
    p & :: & \{ \nu: \emph{double} \ | \ 0 \leq \nu \leq 1 \} \\
    \emph{sim} &::& \mathbb{S} \times \mathbb{S} \to p \\
    \emph{P}_{\bowtie} &::& \emph{Column} \times \emph{Column} \to p \\
    \emph{P}_{\phi} &::& \emph{Column} \times \emph{Value} \to p \\
\end{array}
\]
\vspace{-0.2in}
\caption{Symbols used in sketch completion}
\label{fig:symb-synth}
\end{figure}

%% file: rules-view.tex
\begin{figure}
\centering
\[
\begin{array}{cr}

\irule
{\begin{array}{c}
    t \in \emph{dom}(\Gamma) \\
    p = \emph{sim}(h, t)
\end{array}}
{\Gamma \vdash ??h \Downarrow t : \Gamma(t),~p} & ({\rm Table}) \\ \ \\

\irule
{\begin{array}{c}
    \Gamma \vdash \chi \Downarrow T : \tau,~p_1 \\
    \Gamma, \tau \vdashs \kappa \Downarrow L : \tau_1,~p_2
\end{array}}
{\Gamma \vdash \Pi_\kappa(\chi) \Downarrow \Pi_L(T) : \tau_1,~p_1 \otimes p_2} & ({\rm {Proj}}) \\ \ \\

\irule
{\begin{array}{c}
    \Gamma \vdash \chi \Downarrow T : \tau,~p_1 \\
    \Gamma, \tau \vdashs \psi \Downarrow \phi : \mathbb{B},~p_2
\end{array}}
{\Gamma \vdash \sigma_\psi(\chi) \Downarrow \sigma_{\phi}(T) : \tau,~p_1 \otimes p_2} & ({\rm {Sel}}) \\ \ \\

\irule
{\begin{array}{c}
    \Gamma \vdash \chi_1 \Downarrow T_1 : \tau_1,~p_1 \\
    \Gamma \vdash \chi_2 \Downarrow T_2 : \tau_2,~p_2 \\
    \Gamma, \tau_1 \vdashs ?h_1 \Downarrow c_1 : \{(c_1, \beta)\},~p_3 \\
    \Gamma, \tau_2 \vdashs ?h_2 \Downarrow c_2 : \{(c_2, \beta)\},~p_4 \\
    p = p_1 \otimes p_2 \otimes p_3 \otimes p_4 \otimes \emph{P}_{\bowtie}(c_1, c_2)
\end{array}}
{\Gamma \vdash \join{\chi_1}{?h_1}{?h_2}{\chi_2} \Downarrow \join{T_1}{c_1}{c_2}{T_2} : \tau_1 \cup \tau_2,~p} & ({\rm Join})

\end{array}
\]
\caption{Inference rules for relations}
\label{fig:rules-view}
\end{figure}

%% file: rules-spec.tex
\begin{figure}
\centering
\[
\begin{array}{cr}

\irule
{\begin{array}{c}
    (c, \beta) \in \tau \\
    p = \emph{sim}(h, c)
\end{array}}
{\Gamma, \tau \vdashs ?h \Downarrow c : \{(c, \beta)\},~p} & ({\rm Col}) \\ \ \\

\irule
{\begin{array}{c}
    \Gamma, \tau \vdashs ?h \Downarrow c : \{(c, \beta)\},~p \\
    \emph{type}(f) = \beta \rightarrow \beta'
\end{array}}
{\Gamma, \tau \vdashs f(?h) \Downarrow f(c) : \{(f(c), \beta')\},~p} & ({\rm Fun}) \\ \ \\

\irule
{\begin{array}{c}
    \Gamma, \tau \vdashs f(?h_1) \Downarrow f(c_1) : \{(f(c_1), \beta)\},~p_1 \\
    \Gamma, \tau \vdashs ?h_2 \Downarrow c_2 : \{(c_2, \tau_2)\},~p_2
\end{array}}
{\begin{array}{cc}
    \Gamma, \tau \vdashs & g(f(?h_1), ?h_2) \Downarrow g(f(c_1), c_2) \\
                & : \{(c_2, \tau_2), (f(c_1), \beta)\},~p_1 \otimes p_2
\end{array}} & ({\rm Group}) \\ \ \\

\irule
{\begin{array}{c}
    \Gamma, \tau \vdashs \kappa_1 \Downarrow L_1 : \tau_1,~p_1 \\
    \Gamma, \tau \vdashs \kappa_2 \Downarrow L_2 : \tau_2,~p_2
\end{array}}
{\Gamma, \tau \vdashs \kappa_1, \kappa_2 \Downarrow L_1, L_2 : \tau_1 \cup \tau_2,~p_1 \otimes p_2} & ({\rm ColList}) \\ \ \\

\irule
{\begin{array}{c}
    \Gamma, \tau \vdashs ?h \Downarrow c : \{(c, \beta)\},~p_1 \\
    \Gamma, \tau \vdashs \eta \Downarrow E : \{(c', \beta)\},~p_2 \\
    p = p_1 \otimes p_2 \otimes  \emph{P}_{\phi}(c~op~E)
\end{array}}
{\Gamma, \tau \vdashs ?h~op~\eta \Downarrow c~op~E : \mathbb{B},~p} & ({\rm Pred}) \\ \ \\

\irule
{\begin{array}{c}
     \emph{type}(v) = \beta, \ \  (c, \beta) \in \tau
\end{array}}
{\Gamma, \tau \vdashs v \Downarrow v : \{(v, \beta)\},1.0} & ({\rm Lift}) \\ \ \\

\irule
{\begin{array}{c}
    \Gamma \vdash \chi \Downarrow T: \tau, p 
\end{array}}
{\Gamma, \tau \vdashs \chi \Downarrow T : \tau,p} & ({\rm Reduce}) \\ \ \\

\irule
{\begin{array}{c}
    \Gamma, \tau \vdashs \psi_1 \Downarrow \phi_1 : \mathbb{B},~p_1 \\
    \Gamma, \tau \vdashs \psi_2 \Downarrow \phi_2 : \mathbb{B},~p_2
\end{array}}
{\Gamma, \tau \vdashs \psi_1~lop~\psi_2 \Downarrow \phi_1~lop~\phi_2 : \mathbb{B},~p_1 \otimes p_2} & ({\rm PredLop}) \\ \ \\

\irule
{\begin{array}{c}
    \Gamma, \tau \vdashs \psi \Downarrow \phi : \mathbb{B},~p
\end{array}}
{\Gamma, \tau \vdashs \neg \psi \Downarrow \neg \phi : \mathbb{B},~p} & ({\rm PredNeg})

\end{array}
\]
\vspace{-0.2in}
\caption{Inference rules for specifiers}
\label{fig:rules-spec}
\end{figure}

%% file: refinement.tex
\section{Sketch Refinement}\label{sec:repair}

In the previous section, we saw how to generate a ranked list of possible sketch completions using both types (i.e., database schema) as well as contents of the database.  However, in some cases, \system may fail to find any well-typed, high-confidence completion of the sketch. In particular, this situation can arise for two different reasons:

\begin{enumerate} 
\item If the user does not know the underlying data organization, her description will likely be inaccurate or imprecise. In this case, even a perfect semantic parser will generate query sketches that cannot be successfully instantiated with high confidence.
\item Even though the semantic parser generates all possible query skeletons, this set is typically very large, so \system only considers the top $k$ sketches. Hence, if the correct query sketch is not among the top $k$ choices, \system may not try to instantiate the correct sketch.
\end{enumerate} 
As mentioned in Section~\ref{sec:intro}, \system deals with these difficulties by using program repair to \emph{automatically refine}  query sketches generated by the semantic parser. 

\input{algo-synth}

\input{symb-repair}

The high-level structure of \system's synthesize-repair loop is presented in Algorithm~\ref{algo:synth}. Given one of the top $k$ query sketches generated by the semantic parser, \system synthesizes all possible completions $\Theta$ of sketch $\sketch$ using the inference rules presented in Section~\ref{sec:completion}. If any completion $\inst_i$ has a confidence score $\prob_i$ higher than some threshold $\gamma$, it is  added to the result list $\res$ (line 10). However, if there are no completions of $\sketch$ that exceed our minimum confidence threshold, we then try to repair it at most $n$ times (lines~\ref{line:synth-loopstart}--\ref{line:synth-loopend}). At a high level, the repair procedure first performs fault localization on $\sketch$ to identify the root cause $\sketch_i$ of the problem (line~\ref{line:synth-fault}) and repairs $\sketch$ by replacing the sub-sketch $\sketch_i$ with a suitable refinement (line~\ref{line:synth-ref}). In what follows, we describe fault localization and repair in more detail.

\input{algo-localize}

\subsection{Fault localization} The basic idea underlying our fault localization procedure is to find a \emph{smallest subterm} of sketch $\sketch$ that (a) cannot be instantiated with high confidence, and (b) for which a suitable repair exists.  Since we cannot conclusively refute a well-typed completion of the sketch, our fault localization procedure is confidence-driven and considers a sub-sketch $\sketch$ to be \emph{``faulty"} if no instantiation of $\sketch$ meets our confidence threshold.

Our fault localization procedure is presented in more detail in Algorithm~\ref{algo:localize}. The recursive {\sc FaultLocalize} procedure takes as input the database schema $\Gamma$,  a partial sketch $\sketch$, which can be either a relation or a specifier. If $\sketch$ is a specifier, {\sc FaultLocalize} also takes as input a record type $\tau$, which describes the schema for the parent table. (Recall that sketch completion for specifiers requires the parent table $\tau$.) The return value of {\sc FaultLocalize} is either the faulty sub-sketch or null (if $\sketch$ cannot be repaired).

Let us now consider  Algorithm~\ref{algo:localize} in more detail. If $\sketch$ is a relation,  we first recurse down to its subrelations and specifiers (see Figure~\ref{fig:symb-repair}) to identify a smaller subterm that can be repaired (lines 4--13). On the other hand, if $\sketch$ is a specifier (lines 14--17), we then recurse down to its subspecifiers, again to identify a smaller problematic subterm. If we cannot identify any such subterm, we then consider the current partial sketch $\sketch$ as the possible cause of failure. That is, if $\sketch$ cannot be instantiated in a way that meets our confidence threshold, we then check if $\sketch$ can be repaired using one of the rewrite rules from Figure~\ref{fig:rules-repair}. If so, we return $\sketch$ as the problematic subterm (lines 18--20).

One subtle part of the fault localization procedure is the handling of specifiers in lines 12--14. Recall from Section~\ref{sec:completion} that the instantiation of specifiers is dependent on the instantiation of the parent relation. Specifically, when we instantiate a relation such as $\Pi_\kappa(\chi)$, we need to know the type of $\chi$ when we complete $\kappa$. Hence, there is a valid completion of $\Pi_\kappa(\chi)$ if there \emph{exists} a valid completion of $\kappa$ for \emph{some} instantiation of $\chi$. Thus, we can only say that $\omega_i$ (or one of its subterms) is faulty if it is faulty for all possible instantiations of $\chi$ (i.e., $\forall j. \ \omega_{ij}' \neq {\rm null}$).

\input{rules-repair}

\begin{example}

    Consider the query  ``Find the number of papers in VLDB 2010'' from the motivating example in Section~\ref{sec:overview}. The initial sketch generated by semantic parsing is:
\[
    \Pi_{\emph{count(?papers)}}(\sigma_{? = \emph{VLDB 2010}}(\emph{??papers}))
\]
Since there is no high-confidence completion of this sketch, we perform fault localization using Algorithm~\ref{algo:localize}. The innermost hole  {\tt ??papers} can be instantiated with high confidence, so we next consider the subterm {\tt ? = VLDB 2010}. Observe that there is no completion of {\tt ??papers} under which  {\tt ? = VLDB 2010} has a high confidence score because no table in the database contains the entry {\tt "VLDB 2010"}. Hence, fault localization identifies the predicate as {\tt ? = VLDB 2010} as the root cause of failure.

\end{example}

\vspace{0.05in}
\subsection{Repair} Once \system identifies the faulty subpart of the sketch, it then tries to repair it using a database of repair tactics.  Figure~\ref{fig:rules-repair} shows a {representative subset} of our repair tactics in the form of rewrite rules. At a high-level, we try to repair the sketch by adding new predicates, join operators, and columns.

Let us now consider the tactics shown in Figure~\ref{fig:rules-repair} in more detail.
The \emph{AddPred} rule splits a predicate into two parts by introducing a conjunct. In particular, consider a  predicate $?h~op~v$ and suppose that $v$  is a string that contains a common delimiter (e.g., space, apostrophe etc.). In this case, the \emph{AddPred} tactic splits $v$ into two parts $v_1, v_2$ occurring before and after the delimiter and rewrites the predicate as $?h~op~v_1 \land ?h~op~v_2$. For example, we have used this tactic in the motivating example from Section~\ref{sec:overview} when rewriting {\tt ?="VLDB 2010"} as  {\tt ?="VLDB"} and {\tt ?="2010"}.

The next three rules labeled \emph{AddJoin}  introduce additional join operators in selections, projections, and joins. Since users may erroneously assume that some information is stored in a single database table, the \emph{AddJoin} tactics allow us to introduce join operators in cases where the user's English description is imprecise or misleading. Recall that we also use the \emph{AddJoin} tactic in our example from  Section~\ref{sec:overview}.

The next rule labeled \emph{AddFunc} introduces an aggregate function if the hint $h$ in $?h$ contains the name of an aggregate function (e.g., count) or something similar to it.
For instance, consider a hole {\tt ?} with hint \emph{``average~grade"}. Since \emph{``average"} is also an aggregate function, the \emph{AddFunc} rule can be used to rewrite this as  $\emph{avg(?grade)}$.

Finally, the last rule labeled \emph{AddCol} introduces a new column in the predicate $?h~op~v$. Since $v$ may refer to the name of a column rather than a constant string value, the \emph{AddCol} rule allows us to consider this alternative possibility.

%% file: algo-synth.tex
\begin{algorithm}[t]
\caption{Synthesis Algorithm}
\label{algo:synth}
\begin{algorithmic}[1]

\vspace{0.05in}
\Procedure{\sc Synthesize}{$\query$, $\Gamma$}
\vspace{0.05in}
\State {\rm \bf Input:} natural language query $\query$, schema $\Gamma$
\State {\rm \bf Output:} List $\res$ of SQL queries with probabilities
\vspace{0.05in}

\State $\res$ = $\emptyset$; $[\sketch]$ = {\sc Parse}($\query$)
\ForAll{top $k$ ranked $\sketch \in [\sketch]$}
    \Loop{~$n$ times} \label{line:synth-loopstart}
        \State $\Theta$ = {\sc Instantiate}($\sketch$, $\Gamma$, null) 
        \If{{\sc MaxProb}($\Theta$) $> \gamma$}
            \ForAll{$(\inst_i, \prob_i, \tau_i) \in \Theta$}
                \IIf{$\prob_i > \gamma$}{$\res$.add($\inst_i, \prob_i$)}
            \EndFor
            \State \textbf{break}
        \EndIf
        \State $\sketch_i$ = {\sc FaultLocalize}($\sketch$, $\Gamma$, $\emptyset$) \label{line:synth-fault}
        \IIf{$\sketch_i = $ null}{\textbf{break}}
        \State $\sketch$ = $\sketch[${\sc Repair}($\sketch_i$)$/\sketch_i]$ \label{line:synth-ref}
        \EndLoop \label{line:synth-loopend}
\EndFor
\State \Return $\res$

\EndProcedure
\end{algorithmic}
\end{algorithm}

%% file: symb-repair.tex
\begin{figure}
\centering
\[
\begin{array}{r c l}
    \textsc{SubRelations}(\Pi_\kappa(\chi)) &=& \{(\chi, \kappa)\} \\
    \textsc{SubRelations}(\sigma_\psi(\chi)) &=& \{(\chi, \psi)\} \\
    \textsc{SubRelations}(\join{\chi_1}{?h_1}{?h_2}{\chi_2}) &=& \{(\chi_1, ?h_1), (\chi_2, ?h_2)\} \\
    \textsc{SubSpecifiers}(g(f(?h_1), ?h_2)) &=& \{f(?h_1), ?h_2\} \\
    \textsc{SubSpecifiers}(\kappa_1, \kappa_2) &=& \{\kappa_1, \kappa_2\} \\
    \textsc{SubSpecifiers}(?h~op~\eta) &=& \{?h, \eta\} \\
    \textsc{SubSpecifiers}(\psi_1~lop~\psi_2) &=& \{\psi_1, \psi_2\} \\
    \textsc{SubSpecifiers}(\neg \psi) &=& \{\psi\} \\
\end{array}
\]
\vspace{-0.2in}
\caption{Auxiliary functions used in Algorithm~\ref{algo:localize}}
\label{fig:symb-repair}
\end{figure}

%% file: algo-localize.tex
\begin{algorithm}[t]
\caption{Fault Localization Algorithm}
\label{algo:localize}
\begin{algorithmic}[1]

\vspace{0.05in}
\Procedure{\sc FaultLocalize}{$\sketch$, $\Gamma$, $\tau$}
\vspace{0.05in}
\State {\rm \bf Input:} partial sketch $\sketch$, schema $\Gamma$, record type $\tau$
\State {\rm \bf Output:} faulty partial sketch $\sketch'$ or null
\vspace{0.05in}

\If{{\sc isRelation}($\sketch$)} \label{line:rep-recstart}
    \ForAll{$(\chi_i, \omega_i) \in$ {\sc SubRelations}($\sketch$)}
        \State $\chi_i'$ = {\sc FaultLocalize}($\chi_i$, $\Gamma$, null) \label{line:rep-relloc}
        \IIf{$\chi_i' \neq$ null}{\Return $\chi_i'$}
        \State $\Theta_i$ = {\sc Instantiate}($\chi_i$, $\Gamma$, null)
        \ForAll{$(\inst_j$, $\prob_j$, $\tau_j) \in \Theta_i$}
            \State $\omega_{ij}'$ = {\sc FaultLocalize}($\omega_i$, $\Gamma$, $\tau_j$) \label{line:rep-specloc}
        \EndFor
        \If{$\forall j.~\omega_{ij}' \neq$ null}
            \IIf{$\forall j,k.~\omega_{ij}'=\omega_{ik}'$}{\Return $\omega_{i0}' $}
            \IElseIf{{\sc CanRepair}($\omega_i$)}{\Return $\omega_i$}
        \EndIf
    \EndFor
\ElsIf{{\sc isSpecifier}($\sketch$)}
    \ForAll{$\omega_i \in${\sc SubSpecifiers}($\sketch$)}
        \State $\omega_i'$ = {\sc FaultLocalize}($\omega_i$, $\Gamma$, $\tau$) \label{line:rep-omegaloc}
        \IIf{$\omega_i' \neq$ null}{\Return $\omega_i'$}
    \EndFor
\EndIf \label{line:rep-recend}
\vspace{0.1in} 
\State $\Theta$ = {\sc Instantiate}($\sketch$, $\Gamma$, $\tau$) \label{line:rep-inst} 
\If{{\sc MaxProb}($\Theta$) $< \rho$ \textbf{and} {\sc CanRepair}($\sketch$)} \label{line:rep-cond}
    \State \Return $\sketch$
\EndIf
\IElse{\Return null} \label{line:rep-end}

\EndProcedure
\end{algorithmic}
\end{algorithm}

%% file: rules-repair.tex
\begin{figure}
\centering
\[
\begin{array}{cr}

\irule
{\emph{split}(v) = (v_1, v_2), \ \ v_2 \neq \epsilon}
{ ?h~op~v \rightsquigarrow ?h~op~v_1 \land ?h~op~v_2} & ({\rm AddPred}) \\ \ \\

\irule
{}
{ \sigma_\psi(\chi) \rightsquigarrow \sigma_\psi(\join{\chi}{?\epsilon}{?\epsilon}{??\epsilon})} & ({\rm AddJoin1}) \\ \ \\

\irule
{}
{ \Pi_\kappa(\chi) \rightsquigarrow \Pi_\kappa(\join{\chi}{?\epsilon}{?\epsilon}{??\epsilon})} & ({\rm AddJoin2}) \\ \ \\

\irule
{}
{ \join{\chi_1}{?h_1}{?h_2}{\chi_2} \rightsquigarrow \join{\join{\chi_1}{?\epsilon}{?\epsilon}{??\epsilon}}{?\epsilon}{?\epsilon}{\chi_2}} & ({\rm AddJoin3}) \\ \ \\

\irule
{\begin{array}{c}
    \emph{split}(h) = (f', h')  \\ f \in \emph{AggrFunc}, \ \  \emph{sim}(f, f') \geq \delta
\end{array}}
{ ?h \rightsquigarrow f(?h')} & ({\rm AddFunc}) \\ \ \\

\irule
{
}
{ ?h~op~v \rightsquigarrow ?h~op~?v} & ({\rm AddCol}) \\ \ \\

\end{array}
\]
\vspace{-0.2in}
\caption{Repair tactics. Here, \emph{split}($v$)  tokenizes value $v$ using predefined delimiters. Specifically,  \emph{split}($v$) = ($v_1, v_2$) iff $v_1$ occurs before the first occurrence of the delimiter and $v_2$ occurs after. If the delimiter doesn't appear in $v$, then \emph{split}($v$) = ($v, \epsilon$).}
\label{fig:rules-repair}
\end{figure}

%% file: impl.tex
\section{Implementation}\label{sec:impl}

We have implemented the proposed technique as a tool called \system, written in a combination of C++ and Java. As mentioned in Section~\ref{sec:parse}, \system's semantic parser is implemented on top of the {\sc Sempre} framework~\cite{sempre} and uses the Word2Vec~\cite{word2vec} tool for computing similarity between English hints in the sketch and names of database tables and columns.  Our implementation synthesizes SQL queries for the top 5 sketches generated by the semantic parser and repairs each sketch at most 5 times.

\vspace{0.1in}\noindent
{\bf \emph{Training data.}} Recall from Section~\ref{sec:parse} that our semantic parser uses supervised machine learning to optimize the weights used in the likelihood score for each utterance. Towards this goal, we used queries for a mock database accompanying a databases textbook~\cite{dbSystems}. Specifically, this database contains information about the employees, departments, and projects  for a hypothetical company. In order to train the semantic parser, we extracted the English descriptions of 34 queries from the textbook and manually wrote the corresponding sketch for each query. Note that the English descriptions were obtained directly from the textbook, and our sketches are constructed directly from the English description (without manually repairing them with respect to the actual database schema).

\vspace{0.1in}\noindent
{\bf \emph{Optimizations.}} 
While our implementation closely follows the technical presentation in this paper, it performs two important optimizations that we have not mentioned previously: First, we memoize the result of sketch instantiation for every subterm. Since the fault localization procedure instantiates the same sub-sketch many times, memoizing sketch completions is important for a practical implementation. Second, the implementation of {\sc Instantiate} (recall Algorithm~\ref{algo:synth}) does not generate all possible instantiations of a sketch. In particular, if the score for a subterm is less than a certain confidence threshold, we do not try to instantiate the remaining holes in the sketch. 

\vspace{0.1in}\noindent
{\bf \emph{Confidence thresholds.}} 
Recall from Section~\ref{sec:repair} that our sketch refinement algorithm uses two different confidence thresholds, namely a parameter $\gamma$ from Algorithm~\ref{algo:synth}, which is used to determine if a sketch should be repaired, and a parameter $\rho$ from Algorithm~\ref{algo:localize}, which is used to determine whether a sub-sketch is faulty. In our current implementation, these parameters are hard-coded as $\gamma = 0.35$ and $\rho = 0.45$. These values were manually determined by experimenting with different choices of these parameters  on our training queries obtained from a databases textbook~\cite{dbSystems}.

%% file: evaluation.tex
\section{Evaluation}\label{sec:eval}

To evaluate \system, we perform experiments that are designed to answer the following questions:
\begin{itemize}
    \item[{\bf Q1.}] How effective is \system at synthesizing SQL queries from natural language descriptions? 
    \item[{\bf Q2.}] What is \system's average running time per query? 
    \item[{\bf Q3.}] How well does \system perform across different databases?
    \item[{\bf Q4.}] How does \system perform compared to other state-of-the-art NLIDB systems?
    \item[{\bf Q5.}] How do the use of types, database contents, and repair contribute to the effectiveness of \system in practice?    
\end{itemize}

 \subsection{Experimental Setup}

To answer these research  questions, we evaluate \system on three real-world databases, namely  the Microsoft academic search database (MAS) used for evaluating \nalir~\cite{nalir}, the Yelp business reviewing database (YELP), and the IMDB movie database (IMDB).  Table~\ref{tab:dbstat} provides statistics about each database.

\input{tab-dbstat}

\vspace{0.1in}\noindent
{\bf \emph{Benchmarks.}}
To evaluate \system on these databases, we collected a total of 455 natural language queries.
 For the MAS database, we use the same 196 benchmarks that are used for evaluating {\sc Nalir}~\cite{nalir}. For the IMDB and YELP databases, we asked a group of people at our organization to come up with English queries that they might like to answer using IMDB and YELP.
Participants were given information about the type of data available in each database (e.g., business names, cities, etc.), but they did not have any knowledge of the underlying database schema. Specifically, the participants did not know about the names of database tables, columns, primary and foreign keys etc.

\vspace{0.1in}\noindent
{\bf \emph{Checking correctness.}} In our evaluation, we need to assess whether the queries returned by \system are correct. Since there are many equivalent ways of writing the same query, we manually inspected the results produced by \system to determine which of the queries faithfully implement the user's description with respect to the underlying database schema.

\input{tab-group}

\input{tab-result}

\vspace{0.1in}\noindent
{\bf \emph{Categorization of benchmarks.}}
To assess how \system performs on different classes of queries,  we manually categorize the benchmarks into four  groups based on the characteristics of their corresponding SQL query. Table~\ref{tab:group} shows our taxonomy and provides an English description for each category. While there is no universal agreement on the difficulty level of a given database query, we believe that  benchmarks in category $C_{i+1}$ are generally harder for humans to write than benchmarks in category $C_i$.

\vspace{0.1in}\noindent
{\bf \emph{Hardware and OS.}} 
 All of our experiments are conducted on an Intel Xeon(R) computer with E5-1620 v3 CPU and 32GB memory, running the Ubuntu 14.04 operating system.

\subsection{Accuracy and Running Time}\label{sec:eval-accuracy}

Table~\ref{tab:result} summarizes the results of evaluating \system on the 455 benchmarks involving the MAS, IMDB, and YELP databases. In this table, the column labeled {``Count"} shows the number of benchmarks under each category. The columns labeled {``Top $k$"} show the number (\#) and percentage (\%) of benchmarks whose target query is ranked within the top $k$ queries synthesized by \system. Finally, the columns labeled ``Parse time" and ``Synth/repair time" show the average time (in seconds) for semantic parsing and sketch completion/refinement respectively. 

As shown in Table~\ref{tab:result}, \system reaches close to 90\% accuracy across all three databases when we consider a benchmark to be successful if the desired query appears within the top 5 results. Even if we adopt a stricter definition of success and consider \system to be successful if the target query is ranked within the top one (resp. top three) results, \system still achieves approximately 78\% (resp. 86\%) accuracy. Also, observe that \system's synthesis time is quite reasonable;  on average, \system takes 1.19 seconds to synthesize each query, with 86\% of synthesis time dominated by semantic parsing.

\subsection{Comparison with NALIR}

To evaluate whether the results described in Section~\ref{sec:eval-accuracy} improve over the state-of-the-art, we also compare \system against \nalir, a recent NLIDB system  described in VLDB'14~\cite{nalir}. Similar to \system, the \nalir system also aims to be database-agnostic   (i.e., does not require database-specific training). However, unlike \system, which is fully automated,   \nalir can also be used in an interactive setting that allows the  user to provide  guidance by choosing the right query structure or the names of database elements. In order to perform a fair comparison between \system and \nalir, we use \nalir in the non-interactive setting. Furthermore, since  \nalir only generates a single database query as its output, we compare \nalir's results with the \emph{top-ranked query} produced by \system.

As shown in Figure~\ref{fig:comp}, \system outperforms \nalir on all three databases with respect to the number of benchmarks that can be solved. In particular, \system's average accuracy is 78\% whereas \nalir's average accuracy is less than 32\%. Observe that the queries for the MAS database are the same ones used for evaluating \nalir, but \system outperforms \nalir even on this dataset. Furthermore, even though \system's accuracy is roughly the same across all three databases, \nalir performs significantly worse on the IMDB and YELP databases.

\subsection{Variations of Sqlizer}
In order to evaluate the relative importance of each of component in \system's design, we also compare \system against versions of itself where certain features are disabled. Specifically, we consider  three variants  of \system:
\begin{itemize}
\item {\bf NoData.} In this variant, we do not use the contents of the database in assigning scores during sketch completion. Specifically, we still use the database schema in assigning scores to each possible completion, but we do not use the database contents.
\item {\bf NoType.} In this variant of \system, we do not use any type information in our sketch completion rules. In other words, this version of \system uses the contents of the database, but not the database schema. 
\item {\bf NoRepair.} This final variant of \system does not refine the initial query sketch generated using semantic parsing. In other words, this version does not perform either fault localization or sketch repair.
\end{itemize}

Figure~\ref{fig:SelfComp} shows the results of our evaluation comparing \system against these three variants of itself. As we can see from this figure, disabling any of these three features (namely, type information, database contents, and repair) dramatically reduces the accuracy of the system. In particular, while the full \system system ranks the target query within the top 5 results in over 85\% of the cases, the average accuracy of all three variants is below 50\%.  We believe that these results demonstrate that the use of types, database contents, and sketch refinement are all crucial for the overall effectiveness of \system.

\input{fig-comp}
\input{fig-feature-impact}

%% file: tab-dbstat.tex
\begin{table}[t]
\centering
\begin{tabular}{| c | c | c | c |}
    \hline
    Database & Size & \#Tables & \#Columns \\
    \hline
    MAS & 3.2GB & 17 & 53 \\
    \hline
    IMDB & 1.3GB & 16 & 65 \\
    \hline
    YELP & 2.0GB & 7 & 38 \\
    \hline
\end{tabular}
\vspace{-0.1in}
\caption{Database Statistics}
\label{tab:dbstat}
\end{table}

%% file: tab-group.tex
\begin{table}[t]
\centering
\newcolumntype{A}{ >{\arraybackslash} m{0.3\textwidth} }
\begin{tabular}{| c | A |}
    \hline
    Cat  & \makecell{\centering Description} \\
    \hline  \hline
    C1  & Does not use aggregate function or join operator\\
    \hline
    C2  & Does not use aggregate function but Joins different tables \\
    \hline
    C3  & Uses an aggregate function \\
    \hline
    C4 & Uses subquery or self-join \\
    \hline
\end{tabular}
\vspace{-0.1in}
\caption{Categorization of different benchmarks}
\label{tab:group}
\end{table}

%% file: tab-result.tex
\begin{table*}[t]
\centering
\newcolumntype{A}{ >{\centering\arraybackslash} m{0.055\textwidth} }
\newcolumntype{B}{ >{\centering\arraybackslash} m{0.12\textwidth} }
\begin{tabular}{|c|c|c|A|A|A|A|A|A|B|B|}
\hline
\multirow{2}{*}{{\bf DB}} & \multirow{2}{*}{{\bf Cat}} & \multirow{2}{*}{{\bf Count}} & \multicolumn{2}{c|}{{\bf Top 1}} & \multicolumn{2}{c|}{{\bf Top 3}} & \multicolumn{2}{c|}{{\bf Top 5}} & {\bf Parse} & {\bf  Synth/repair} \\
\cline{4-9}
 & & & {\bf \#} & {\bf \%} & {\bf \#} & {\bf \%} & {\bf \#} & {\bf \%} & {\bf time (s)} & {\bf  time (s)} \\
\hline \hline
\multirow{5}{*}{\begin{turn}{90}\makecell{{\bf MAS}}\end{turn}}
 & C1 & 14 & 12 & 85.7 & 14 & 100.0 & 14 & 100.0 & 0.38 & 0.07 \\
\cline{2-11}
 & C2 & 59 & 52 & 88.1 & 54 & 91.5 & 54 & 91.5 & 1.04 & 0.14 \\
\cline{2-11}
 & C3 & 60 & 48 & 80.0 & 54 & 90.0 & 54 & 90.0 & 1.05 & 0.24 \\
\cline{2-11}
 & C4 & 63 & 45 & 71.4 & 47 & 74.6 & 51 & 81.0 & 2.39 & 0.16 \\
\cline{2-11}
 & {\bf Total} & {\bf 196} & {\bf 157} & {\bf 80.1} & {\bf 169} & {\bf 86.2} & {\bf 173} & {\bf 88.3} & {\bf 1.43} & {\bf 0.17} \\
\cline{2-11}
\hline \hline
\multirow{5}{*}{\begin{turn}{90}\makecell{{\bf IMDB}}\end{turn}}
 & C1 & 18 & 16 & 88.9 & 17 & 94.4 & 17 & 94.4 & 0.47 & 0.16 \\
\cline{2-11}
 & C2 & 69 & 49 & 71.0 & 57 & 82.6 & 59 & 85.5 & 0.62 & 0.24 \\
\cline{2-11}
 & C3 & 27 & 24 & 88.8 & 26 & 96.2 & 26 & 96.2 & 0.72 & 0.36 \\
\cline{2-11}
 & C4 & 17 & 11 & 64.7 & 12 & 70.5 & 13 & 76.4 & 0.67 & 0.33 \\
\cline{2-11}
 & {\bf Total} & {\bf 131} & {\bf 100} & {\bf 76.3} & {\bf 112} & {\bf 85.4} & {\bf 115} & {\bf 87.7} & {\bf 0.62} & {\bf 0.26} \\
\cline{2-11}
\hline \hline
\multirow{5}{*}{\begin{turn}{90}\makecell{{\bf YELP}}\end{turn}}
 & C1 & 8 & 6 & 75.0 & 7 & 87.5 & 7 & 87.5 & 0.52 & 0.02 \\
\cline{2-11}
 & C2 & 49 & 38 & 77.5 & 40 & 81.6 & 43 & 87.7 & 0.71 & 0.05 \\
\cline{2-11}
 & C3 & 51 & 41 & 80.3 & 48 & 94.1 & 49 & 96.0 & 0.69 & 0.04 \\
\cline{2-11}
 & C4 & 20 & 15 & 75.0 & 15 & 75.0 & 15 & 75.0 & 0.90 & 0.06 \\
\cline{2-11}
 & {\bf Total} & {\bf 128} & {\bf 100} & {\bf 78.1} & {\bf 110} & {\bf 85.9} & {\bf 114} & {\bf 89.0} & {\bf 0.85} & {\bf 0.05} \\
\cline{2-11}
\hline
\end{tabular}
\vspace{-0.1in}
\caption{Summary of our experimental evaluation}
\label{tab:result}
\end{table*}

%% file: fig-comp.tex
\definecolor{bblue}{HTML}{0064FF}
\definecolor{rred}{HTML}{C0504D}
\definecolor{ggreen}{HTML}{9BBB59}
\definecolor{ppurple}{HTML}{9F4C7C}

\begin{figure}[!t]
\centering

\begin{tikzpicture}
  \begin{axis}[
    width = 8cm,
    height = 8cm,
    y = 0.05cm,
    x = 2.4cm,
    major x tick style = transparent,
    ybar = 5*\pgflinewidth,
    bar width = 12pt,
    ymajorgrids = true,
    ylabel = {Percertage Solved (\%)},
    ylabel style = {yshift=-3mm},
    symbolic x coords = {MAS,IMDB,YELP},
    xtick = data,
    scaled y ticks = false,
    enlarge x limits = 0.25,
    ymin = 0,
    ymax = 100,
    legend cell align = left,
    legend style = {
      at = {(0.5, 1.08)},
      legend columns = -1,
      anchor = north,
    }
  ]
  \hspace*{-2mm}
    \addplot[style={bblue,fill=bblue,mark=none}]
    coordinates {(MAS,80.1)(IMDB,76.3)(YELP,78.2)};

    \addplot[style={ggreen,fill=ggreen,mark=none}]
    coordinates {(MAS,54.1)(IMDB,16.8)(YELP,10.9)};

    \legend{\system,\nalir}
  \end{axis}
\end{tikzpicture}
\vspace{-0.1in}
\caption{Comparison between \system and \nalir}
\label{fig:comp}
\end{figure}

%% file: fig-feature-impact.tex
\definecolor{bblue}{HTML}{0064FF}
\definecolor{rred}{HTML}{C0504D}
\definecolor{ggreen}{HTML}{9BBB59}
\definecolor{ppurple}{HTML}{9F4C7C}

\begin{figure}
\centering

\begin{tikzpicture}
  \begin{axis}[
    width = 8cm,
    height = 8cm,
    y = 0.05cm,
    x = 2.4cm,
    major x tick style = transparent,
    ybar = 5*\pgflinewidth,
    bar width = 10pt,
    ymajorgrids = true,
    ylabel = {Percentage Solved (\%)},
    ylabel style = {yshift=-4mm},
    symbolic x coords = {MAS,IMDB,YELP},
    xtick = data,
    scaled y ticks = false,
    enlarge x limits = 0.25,
    ymin = 0,
    ymax = 100,
    legend cell align = left,
    legend style = {
      at = {(0.5, 1.08)},
      legend columns = -1,
      anchor = north,
    }
  ]
  \hspace*{-2mm}
    \addplot[style={bblue,fill=bblue,mark=none}]
    coordinates {(MAS,88.2)(IMDB,87.7)(YELP,89.0)};

    \addplot[style={rred,fill=rred,mark=none}]
    coordinates {(MAS,35.2)(IMDB,41.9)(YELP,17.1)};

    \addplot[style={ggreen,fill=ggreen,mark=none}]
    coordinates {(MAS,33.6)(IMDB,14.5)(YELP,28.9)};

    \addplot[style={ppurple,fill=ppurple,mark=none}]
    coordinates {(MAS,20.9)(IMDB,22.1)(YELP,19.5)};

    \legend{\system,NoData,NoType,NoRepair}
  \end{axis}
\end{tikzpicture}
\vspace{-0.1in}
\caption{Comparison between different variations of \system on Top 5 results}
\label{fig:SelfComp}
\end{figure}
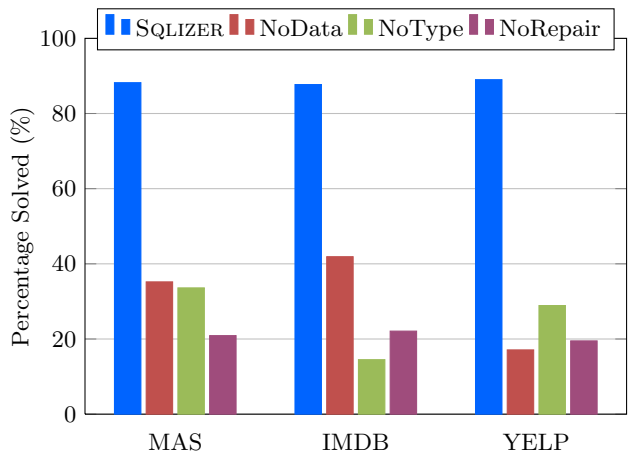

%% file: related.tex
\section{Related Work}\label{sec:related}

The work presented in this paper spans a variety of different topics from the databases, programming languages, software engineering, and natural language processing communities. In what follows, we explain how our approach relates to prior work from these communities.

\vspace{0.1in}\noindent
{\bf \emph{Keyword search.}}
Keyword-based interfaces to databases allow non-experts to formulate ad-hoc queries using a set of keywords or very simple English sentences. Early examples of work in this area include  systems such as BANKS~\cite{banks}, DISCOVER~\cite{discover}, and DBExplorer~\cite{dbexplorer}. More recent work in this area~\cite{soda,sqak,precis} can also support aggregate functions~\cite{sqak} or logical connectives in predicates~\cite{precis}. Generally speaking, \system (and other NLIDB systems) support a richer class of queries than keyword-based systems.

\vspace{0.1in}\noindent
{\bf \emph{Natural language interface to databases (NLIDB).}}
In contrast to other database query interfaces, such as keyword-based or form-based ones, NLIDBs can allow users without any technical skills to formulate more complex database queries. Hence, NLIDBs have been the subject of decades of research ~\cite{nlidb-survey1}. Early work in this area focuses on systems that are hand-crafted to specific databases~\cite{lunar,nlidb3}. Later work describes NLIDB systems that can be reused for multiple databases with  appropriate customization~\cite{team2,mooney1,mooney3}. However, these techniques are not  database-agnostic in that they require additional customization  for each database. In contrast, the technique we describe in this paper is completely database-agnostic and does not require database-specific training. 
Similar to our proposed approach, recent NLIDB systems such as {\sc Precise}~\cite{precise, precise2} and \nalir~\cite{nalir} also aim to be database-agnostic.

The {\sc Precise} system is based on the intuition that certain classes of English descriptions can always be accurately mapped to an unambiguous SQL query. Based on this insight, Popescu et al. consider a class of natural language queries, called \emph{semantically traceable}  natural language descriptions, and describe a system for translating such queries to SQL~\cite{precise,precise2}. While the {\sc Precise} system achieves 100\% accuracy for queries that are semantically traceable, queries that fall outside of this class are rejected by {\sc Precise}.

In contrast to {\sc Precise}, the more recent {\sc Nalir} system does not restrict the class of queries that can be answered and proposes user interaction to achieve reliability.  Specifically, \nalir leverages an English dependency parser to generate linguistic parse trees, which are subsequently translated into \emph{query trees}, possibly with guidance from the user. In addition to being relatively easy to convert to SQL, these query trees can also be translated back into natural language with the goal of facilitating user interaction. In contrast, our goal in this paper is to develop a system that is as reliable as \nalir without requiring guidance from the user. In particular, since users may not be familiar with the underlying database schema, it may be difficult for them to answer some of the questions posed by a \nalir-style system. In contrast, our approach does not  assume that users are familiar with the underlying organization of the data.

\vspace{0.1in}\noindent
{\bf \emph{Query synthesis from other specificions.}}
There is also a significant body of work on automatically synthesizing database queries from informal specifications other than natural language. In one line of work on query synthesis,  users convey their intent   through the use of \emph{input-output examples}~\cite{sqlsynthesizer,qbo,vldb75}. Specifically, an example  consists of a pair of input, output tables where the input tables represent  a miniature version of the database, and the output table is the data that should be extracted using the target query.
We believe that natural language specifications are more user-friendly compared to input-output examples for two  reasons:
First, in order to provide input-output examples, the user must be familiar with the database schema, which is not always the case. Second, since a database may contain several tables with many different columns, providing input-output examples may be prohibitively cumbersome.

Another line of work on query synthesis generates more efficient SQL queries from code written in conventional programming languages~\cite{oopsla08, qbs}. For instance, QBS~\cite{qbs} transforms parts of the application logic into SQL queries by automatically inferring loop invariants. This line of work is not tailored towards end-users and can be viewed as a form of query optimization using static analysis of source code.

\vspace{0.1in}\noindent
{\bf \emph{Programming by natural language.}} Since SQL is a declarative  language, the techniques proposed in this paper are also related to \emph{programming by natural language}~\cite{nlyze,icse16,ijcai15,langToCode,smartsynth}. In addition to query synthesis, natural language has also been used  in the context of smartphone automation scripts~\cite{smartsynth}, ``if-then-else recipes"~\cite{langToCode}, spreadsheet programming~\cite{nlyze}, and string manipulation~\cite{ijcai15}. Among these systems, the {\sc NLyze} tool~\cite{nlyze} is most closely related to \system in that it also combines semantic parsing with type-directed synthesis.  However, {\sc NLyze} does not generate program sketches and uses type-based synthesis to mitigate the low recall of semantic parsing. In contrast, \system uses semantic parsing to generate an initial query sketch, which is refined using repair tactics and completed using type- and content-driven synthesis. Furthermore, {\sc NLyze} targets spreadsheets rather than relational databases.

\vspace{0.1in}\noindent
{\bf \emph{{Program synthesis.}}} The techniques proposed in this paper borrow insights from recent synthesis work in the programming languages community. In particular, the use of the term \emph{query sketch} is inspired by the {\sc Sketch} system~\cite{sketch3} in which the user writes a \emph{program sketch} containing holes (unknown expressions). However, in contrast to the {\sc Sketch} system where the holes are instantiated with constants, holes in our query sketches are completed using tables, columns, and predicates. Similar to \system, some prior techniques (e.g.,~\cite{sqlsynthesizer,sypet}) have also  decomposed the synthesis task into two separate sketch generation and sketch completion phases. However, to the best of our knowledge, we are the first to generate program sketches from natural language using semantic parsing.

In addition to program sketching, this paper also draws insights from recent work on \emph{type-directed program synthesis}. For instance, $\lambda^2$~\cite{lambda2} and {\sc Myth}~\cite{myth} both use type information to drive synthesis.  However, while these techniques synthesize recursive functions from input-output examples or refinement types, we synthesize database queries from natural language. Furthermore, in addition to using type information, our synthesis approach also leverages the contents of the database to rank the results.

\vspace{0.1in}\noindent
{\bf \emph{{Fault localization and program repair.}}} The sketch refinement strategy in \system is inspired by recent work on \emph{fault localization}~\cite{repair-maxsat1} and \emph{program repair}~\cite{spr,prophet,genetic1,semfix}. Similar to other techniques on program repair, the repair tactics employed by \system can be viewed as a pre-defined set of templates for mutating the program. However, to the best of our knowledge, we are the first to apply repair at the level of program sketches rather than programs. 

Fault localization techniques~\cite{repair-maxsat1} aim to pinpoint a program expression that corresponds to the root cause of a bug. Similar to these fault localization techniques, \system tries to identify the smallest problematic subpart of the ``program". However, we perform fault localization at the level of program sketches by finding a smallest sub-sketch for which no high-confidence completion exists.

\vspace{0.1in}\noindent
{\bf \emph{{Semantic parsing.}}}
Unlike syntactic parsing which focuses on the grammatical divisions of a sentence, semantic parsing represents a sentence through logical forms expressed in some formal language~\cite{sempre,sem-parse1,semantic-parsing1,mooney2,mooney3}. Previous techniques have used semantic parsing to directly translate English sentences to queries~\cite{atis,mooney1,mooney2,mooney3}. In contrast, we only use semantic parsing to generate an initial query sketch rather than the full query. We believe that there are two key advantages to our approach: First, our technique can be used to answer queries on a database on which it has not been previously trained. Second, the use of sketch refinement allows us to handle situations where the user's description does not accurately reflect the underlying database schema.

%% file: conclusion.tex
\section{Conclusions}\label{sec:conclusion}

We have proposed a novel, database-agnostic technique for synthesizing SQL queries from natural language. Our approach uses semantic parsing to generate an initial query sketch, which is then completed using type-directed program synthesis and potentially refined using repair tactics. Our empirical evaluation shows that such a marriage of ideas from the databases, programming languages, and natural language processing communities is quite effective: Our tool, \system, performs quite well across multiple databases and outperforms  a state-of-the-art NLIDB system.